\documentclass[runningheads]{llncs}
\usepackage[T1]{fontenc}
\usepackage{amsmath,amssymb,amsfonts}
\usepackage{graphicx,verbatim}
\usepackage{multirow}
\usepackage[hidelinks]{hyperref}
\begin{document}
\title{AtlasGS: Brain MRI Spatial Resolution Harmonization With Shared Gaussian Geometry}
\titlerunning{Gaussian Geometry for MRI Harmonization}
%

\author{Yifan Gao\inst{1,2} \and
Peiran Xu\inst{1,3} \and
Yimeng He\inst{1,4} \and 
Haoran Li\inst{1,3} \and 
Ziyang Long\inst{1,3} \and
Yufeng Wang\inst{1} \and 
Ju Dong Yang\inst{1,5} \and
Debiao Li\inst{1,3}}
%
\authorrunning{Y. Gao et al.}
%
\institute{Biomedical Imaging Research Institute, Cedars-Sinai Medical Center, Los Angeles, CA, USA\\
\email{\{Yifan.Gao, Peiran.Xu, Yimeng.He, Haoran.Li, Ziyang.Long, Yufeng.Wang\}@cshs.org, \{JuDong.Yang, Debiao.Li\}@csmc.edu}
\and Tsinghua Medicine, Tsinghua University, Beijing, China \and Department of Bioengineering, University of
California-Los Angeles, Los
Angeles, CA, USA \and Department of Computational Medicine, Cedars-Sinai Medical Center, Los Angeles, CA, USA \and Karsh Division of Gastroenterology and Hepatology, Cedars-Sinai Medical Center, Los Angeles, CA, USA
\\
} 

\maketitle              
\begin{abstract}
\begin{sloppypar}
Clinical magnetic resonance imaging (MRI) protocols are spatially heterogeneous across modalities and clinical objectives, compromising multimodal joint analysis and arbitrary-view generation. To harmonize MRI resolution, we leverage the commonly acquired isotropic T1-weighted sequence in neuroimaging protocols. We introduced AtlasGS, a Gaussian Splatting (GS)-based shared geometry framework adopts a two-stage training strategy, in which an explicit, subject-specific Gaussian scaffold encoding anatomical geometry is first learned from the isotropic structural scan and then reused to fit appearance for target modalities acquired with sparse slices. Experiments on the UK Biobank, GBM, and ABCD datasets for through-plane super-resolution across multiple modalities (T2-weighted, FLAIR, DWI, ASL), degradation factors ($\times 3$, $\times 5$, $\times 7$), and pathological abnormalities (glioblastoma) demonstrate state-of-the-art reconstruction fidelity. The shared Gaussian geometry enables arbitrary-view generation for target modalities with strong structural consistency and further shows potential for self-supervised in-plane super-resolution. This work establishes explicit geometry-guided representations as a novel, flexible, and interpretable pathway toward retrospective multi-contrast MRI harmonization and reliable clinical reference construction. Source code is available at: \href{https://github.com/yfgao76/AtlasGS}{\texttt{https://github.com/yfgao76/AtlasGS}}.
\end{sloppypar}

\keywords{Super-resolution \and Multi-contrast MRI \and Gaussian Splatting.}

\end{abstract}

\begin{figure}[t]
\centering
\includegraphics[width=\textwidth]{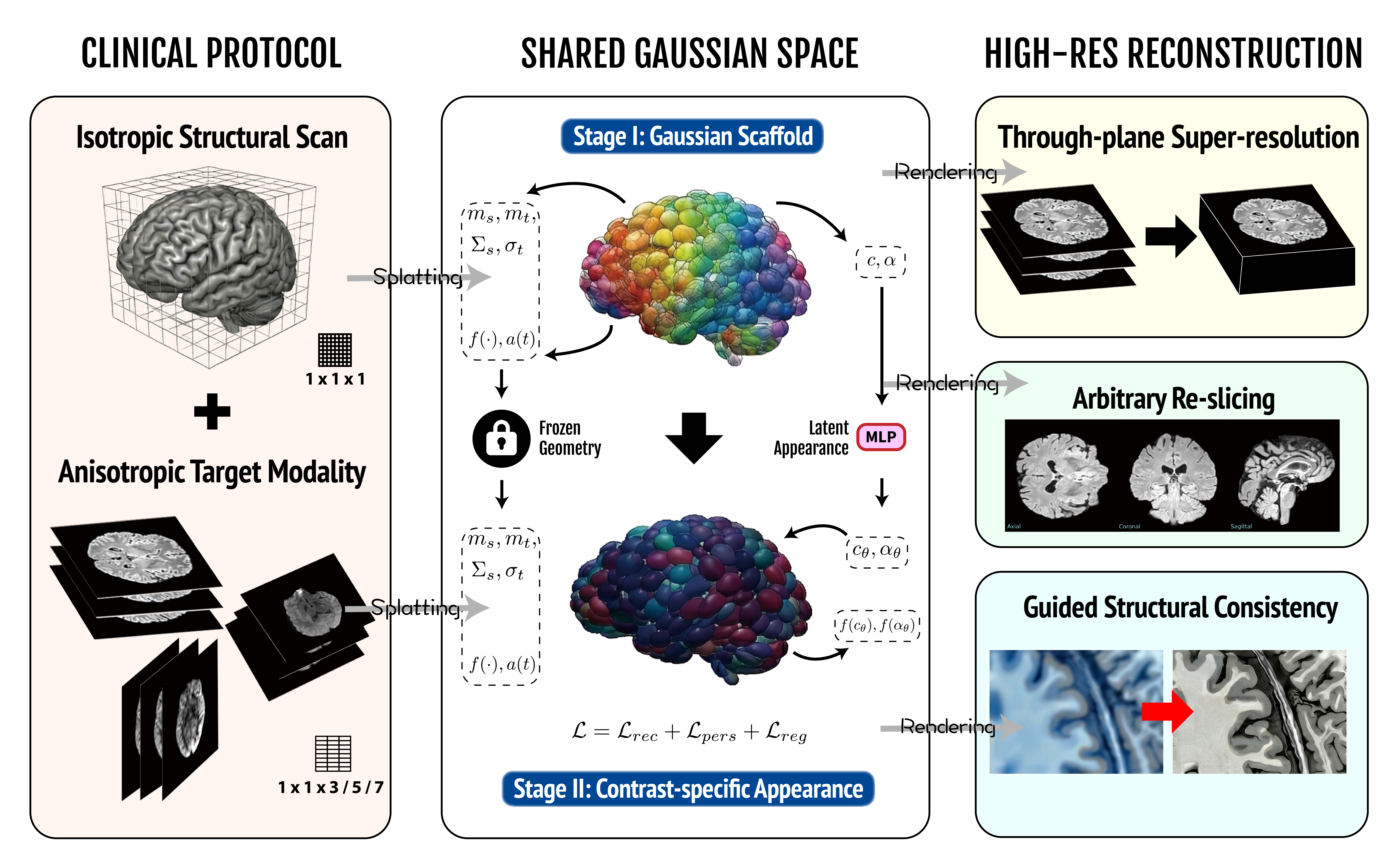}
\caption{Overview of AtlasGS: shared Gaussian geometry framework for multi-contrast MRI spatial resolution harmonization.}
\label{fig1}
\end{figure}

\section{Introduction}

Magnetic resonance imaging (MRI) spatial harmonization refers to establishing a consistent anatomical representation across heterogeneous acquisition protocols, enabling reliable cross-contrast visualization, comparison, and measurement \cite{ref_hu2023harmonization,ref_abbasi2024dlharm,ref_yang2025mriharm_survey}. In this paper, we focus on clinical multi-contrast neuroimaging, where protocols comprise a high-resolution isotropic structural scan together with multiple disease-targeted sequences acquired using anisotropic slice thickness and modality-specific resolutions, trajectories, and acquisition parameters \cite{ref_mahmoudzadeh2014_jmi_srr}. This variability results in inconsistent geometric representations across contrasts and limits coherent arbitrary-view reconstruction, particularly in retrospective datasets where acquisition settings are pre-arranged and hard to simulate precisely \cite{ref_greenspan2002_interslice_sr}, while also indicates the potential of utilizing complementary contrast-specific information \cite{ref_liu2021_unregistered_mc,ref_remedios2023sashimi_smore}.

Conventional MRI protocol harmonization approaches face substantial challenges under diverse imaging modalities and sparse through-plane supervision \cite{ref_khateri2025_mri_sr_survey}. Interpolation produces over-smoothed images and does not align geometric representations across contrasts. Deep learning-based super-resolution models rely on external training data and often struggle with clinical datasets that exhibit wide variability and individualized harmonization requirements \cite{ref_choi2025_tesla}. Implicit Neural Representations (INR) and Gaussian Splatting (GS) have recently gained attention due to their zero-shot continuous coordinate modeling and resolution-free rendering capabilities \cite{ref_molaei2023_inr_survey,ref_kerbl2023_3dgs}. However, these methods do not inherently encode semantic organ geometry, and their potential in bridging multiple contrasts remains under-explored.

Building upon GS, we propose \textbf{AtlasGS: shared Gaussian geometry representation} that learns an anatomical prior while accommodating modality-specific signal formation. Following the MedGS framework \cite{ref_marzol2025_medgs}, we apply video Gaussian splatting \cite{ref_smolak2024_vegas} to an isotropic structural scan to construct an explicit Gaussian scaffold. Contrast-specific appearance parameters are then fitted on this fixed scaffold to enable zero-shot through-plane arbitrary super-resolution across different modalities and resolutions. To enhance structural stability, we incorporate topology-aware Gaussian densification \cite{ref_shen2024_topology_gs} and latent appearance regularization, and integrate low-resolution modality consistency to suppress noise and refine contrast fidelity (see Fig. \ref{fig1}).

\begin{sloppypar}
\begin{sloppypar}
Experiments on UK Biobank (healthy older cohorts), GBM (glioblastoma cohorts), and ABCD (healthy younger cohorts) across multiple modalities (T2-weighted, T2-FLAIR, DWI, ASL) and through-plane super-resolution factors ($\times$3, 5, 7) demonstrate state-of-the-art reconstruction fidelity and cross-modality geometric consistency compared to subject-wise interpolation and contrast-agnostic INR/GS baselines, especially in clinical settings where slice thickness often exceeds 5 mm. We further show that our framework mitigates tumor hallucinations commonly observed in interpolation-based reconstructions, and that the learned Gaussian scaffold can also support in-plane super-resolution. These findings support our hypothesis that explicit geometry-guided representations provide a practical pathway toward retrospective multi-contrast MRI harmonization and the construction of reliable clinical reference spaces.
\end{sloppypar}
\end{sloppypar}

\section{Methods}

Let $I^{T1} \in \mathbb{R}^{H \times W \times D}$ denote an isotropic structural MRI (denoted as T1 in this section),
and let $I^{m}_{LR}$ denote a target modality acquired with thick through-plane
sampling (in low resolution). Our objective is to learn a continuous high resolution representation
\begin{equation}
I^{m}_{HR}(s,t), \quad s=(x,y)\in\mathbb{R}^2,\; t\in\mathbb{R},
\end{equation}
that:

(i) preserves structural geometry defined by $I^{T1}$,

(ii) respects target acquisition physics, and 

(iii) enables arbitrary continuous re-slicing.

\subsection{2D+t Gaussian Geometry with Latent Opacity and Appearance}

We utilize a shared 2D+t Gaussian field from \cite{ref_smolak2024_vegas}
\begin{equation}
\mathcal{G} = \{ \mathcal{G}_i \}_{i=1}^{N},
\end{equation}

where each primitive $\mathcal{G}_i$ represents a localized anatomical element and consists of:

\begin{itemize}
\item \textbf{geometric parameters:} spatial center $m_{s,i} \in \mathbb{R}^2$, through-plane center $m_{t,i} \in \mathbb{R}$, spatial covariance $\Sigma_{s,i} \in \mathbb{R}^{2\times2}$, through-plane variance $\sigma_{t,i}^2$,
\item \textbf{modality-specific parameters:} opacity $\alpha_i$, appearance $c_i$,
\item \textbf{deformation modeling parameters:} deformation function $f_i(\cdot)$, variance modulation function $a_i(t)$.
\end{itemize}

Starting from a separable Gaussian in $(s,t)$,
we allow the spatial center to vary smoothly with $t$, thus the Folded-Gaussian density is defined as
\begin{equation}
FN_i(s,t)
=
\mathcal{N}
\big(
m_{s,i}+f_i(m_{t,i}-t),
a_i(t)\Sigma_{s,i}
\big)
\cdot
\mathcal{N}(m_{t,i},\sigma_{t,i}^2).
\end{equation}

This formulation models nonlinear anatomical continuity along the
through-plane axis. 

Conditioning of $\mathcal{G}$ at fixed $t$ produces a standard 2D Gaussian
in $(x,y)$, and rendering is formulated as:

\begin{equation}
I(s,t)
=
\sum_{i=1}^{N} \alpha_i FN_i(s,t)c_i.
\end{equation}

\subsection{Training Objectives}

Training of $\mathcal{G}$ is performed in two stages:

\begin{itemize}
\item \textbf{Stage I (Structural learning):} to learn geometric parameters from $I^{T1}$.
\item \textbf{Stage II (Modality fitting):} to fit modality-specific appearance using $I^{m}_{LR}$ with frozen geometry.
\end{itemize}

In \textbf{Stage I}, all parameters are optimized towards $I^{T1}$, while in \textbf{Stage II}, only $(\alpha_i, c_i)$ are optimized. 

\paragraph{Reconstruction.}
Reconstruction loss is formulated as

\begin{equation}
\mathcal{L}_{rec}
=
(1-\lambda_{ssim})\|\hat{I}-I_{gt}\|_1
+
\lambda_{ssim}(1-SSIM).
\end{equation}

In \textbf{Stage I}, $\hat{I}$ is exactly $I$. While in \textbf{Stage II}, thick-slice supervision is modeled via slab integration along the
through-plane coordinate $t$. For a slice centered at $t_0$ with
thickness $\Delta t$, the predicted low-resolution slice is obtained by
integrating the continuous Gaussian field within the slab:

\begin{equation}
\hat{I}(s, t_0)
=
\frac{1}{K}
\sum_{k=1}^{K}
I\big(s,\, t_0 + \delta_k\big),
\quad
\delta_k \in \left[-\frac{\Delta t}{2}, \frac{\Delta t}{2}\right],
\end{equation}

where $\{\delta_k\}_{k=1}^{K}$ are uniformly spaced offsets
inside the slice thickness.

\paragraph{Topology preservation.}
To suppress structural artifacts and preserve anatomical integrity, following Topology-GS \cite{ref_shen2024_topology_gs}, 
we impose a Persistance Homology (PH)-based constraint \cite{ref_clough2022_topoloss} $\mathcal{L}_{pers}$.

\paragraph{Latent appearance fitting.}
To ensure stable cross-modality fitting,
appearance is factorized through a low-dimensional latent code:

\begin{equation}
c_i = f_\phi(\theta_i),
\end{equation}

where $\theta_i$ is a compact per-Gaussian embedding and
$f_\phi$ is a shared decoder.

Spatial smoothness is enforced by a kNN graph over Gaussian centers \cite{ref_belkin2003_laplacian} and parameter regularization are applied:

\begin{equation}
\mathcal{L}_{reg}
=
\lambda_{smooth}\sum_{(i,j)} \|\theta_i-\theta_j\|_2^2
+
\lambda_\theta\|\theta\|_2^2.
\end{equation}

\paragraph{Total Loss.}
\begin{equation}
\mathcal{L}
=
\mathcal{L}_{rec}
+
\lambda_{pers}\mathcal{L}_{pers}
+
\mathcal{L}_{reg}.
\end{equation}

Two stages share the same loss. Stage I optimizes on $I^{T1}$, and Stage II optimizes on the target modality.

\subsection{Low-Resolution (LR) Consistent Fusion}

Structure-guided reconstruction may introduce anatomical artifacts
under sparse thick-slice supervision. To balance structural prior and signal
fidelity, we apply a LR consistent fusion at final rendering.

For high-resolution $I^{T1}_{HR}$ and $I^{m}_{HR}$ trained respectively from T1 and separately from target modalitiy, and let $\mathcal{D}(\cdot)$ be the slab-aware
degradation operator (blur + average pooling along $t$), we compute a voxel-wise LR confidence map via
\begin{equation}
w_{LR}
=
\mathrm{softmax}_{\{T1,single\}}
\left(
-\frac{1}{\tau}
\left[
\left|\mathcal{D}(I^{T1}_{HR})-I^{m}_{LR}\right|,
\;
\left|\mathcal{D}(I^{m}_{HR})-I^{m}_{LR}\right|
\right]
\right)_{T1},
\end{equation}
where $\tau$ controls sensitivity and $(\cdot)_{T1}$ selects the T1-guided
channel. After smoothing and clipping, $w_{LR}$ is resampled as $w$ and used
to fuse:
\begin{equation}
I_{fused}= w\, I^{T1} + (1-w)\, I^{single}.
\end{equation}

\section{Experiments and Results}
\subsubsection{Datasets}

We adopt \underline{multi-contrast through-plane super-resolution} as an evaluation task of our representation. Experiments were done on three datasets:
UK Biobank (UKBB) \cite{ref_ukbb_sudlow2015}, UPenn-GBM (GBM) \cite{ref_upenn_gbm_bakas2022} and The Adolescent Brain Cognitive Development (ABCD) study \cite{ref_abcd_casey2018_imaging_across21}. We gratefully curated ABCD data from FOMO300K dataset \cite{Cerri2026large}. 4 target MRI sequences besides isotropic T1-MPRAGE are included, two of the them (T2-TSE and FLAIR) with consistent spacing, while the other 2 (diffusion-weighted imaging, DWI, and arterial spin labeling, ASL) have irregular slice spacing. UKBB and ABCD include respectively elder and younger healthy cohort, while GBM additionally includes patients with glioblastoma, where tumors alter brain anatomy (see Table~\ref{tab:datasets} for details). All volumes underwent skull stripping using HD-BET \cite{ref_hdbet_isensee2019}. Low-resolution inputs are synthetically degraded along the slice direction with Gaussian blur averaging neighboring slices. Each method restores degraded frames from the input. We \textbf{don't intentionally normalize} in order to evaluate our method’s ability to preserve absolute signal values, which is essential for downstream extension to quantitative MRI modalities (e.g. T1/T2 mapping).

\renewcommand{\arraystretch}{1.15}

\begin{table}[t]
\centering
\caption{Dataset configuration, target modality resolution (in mm) and shape, and degradation factors.}
\label{tab:datasets}
\fontsize{8}{9}\selectfont
\begin{tabular}{lccccccc}
\hline
\textbf{Dataset} & \textbf{Train} & \textbf{Test} & \textbf{T1 Res.} & \textbf{T1 Shape} & \textbf{Target} & \textbf{Target Res.} & \textbf{Degrade Ratio} \\
\hline

\multirow{1}{*}{UKBB}
& \multirow{1}{*}{100}
& \multirow{1}{*}{100}
& \multirow{1}{*}{1.0$\times$1.0$\times$1.0}
& (172,228,202)
& Flair
& 1.0$\times$1.0$\times$1.0
& $\times$3 / $\times$5 / $\times$7 \\

\hline

\multirow{2}{*}{GBM}
& \multirow{2}{*}{60}
& \multirow{2}{*}{59}
& \multirow{2}{*}{1.0$\times$1.0$\times$1.0}
& \multirow{2}{*}{(240,240,140)}
& T2w
& 1.0$\times$1.0$\times$1.0
& $\times$7 \\
& & & &
& Flair
& 1.0$\times$1.0$\times$1.0
& $\times$7 \\

\hline

\multirow{2}{*}{ABCD}
& \multirow{2}{*}{34}
& \multirow{2}{*}{33}
& \multirow{2}{*}{0.47$\times$0.47$\times$0.8}
& \multirow{2}{*}{(512,512,256)}
& ASL
& 1.875$\times$1.875$\times$3.5
& $\times$3 \\
& & & &
& DWI
& 0.86$\times$0.86$\times$2.2
& $\times$3 \\

\hline
\end{tabular}
\end{table}

\subsubsection{Implementation}

For each subject, training proceeds in three stages: (1) Train T1 Gaussian scaffold; (2) Train single-modality Gaussian model on target modality; (3) Train and fit T1-guided model with frozen geometry on target modality. Each stages run for 10k iterations. Training takes an average of 20 minutes for one subject on an NVIDIA H100 GPU with 80 GB memory.

Optimization uses Adam \cite{ref_adam_kingma2014}. Feature learning rate is $2.5\times10^{-3}$, opacity $2.5\times10^{-2}$, latent appearance $2.5\times10^{-3}$, and decoder $1\times10^{-3}$. Latent dimension is 4. Graph smoothness uses $k=8$ neighbors. For loss regularization, $\lambda_{ssim}=0.2$, $\lambda_{pers}=10$, $\lambda_{smooth}=0.02$, $\lambda_{\theta}=0.001$.

\subsubsection{Comparative Experiments} 

We compared with through-plane super-resolution methods, include two interpolation methods: slab-wise linear interpolation (Linear), cubic B-spline interpolation (Cubic) \cite{ref_bspline_lehmann2001}; two single-modality INR/GS based methods: MedGS \cite{ref_marzol2025_medgs}, SA-INR \cite{ref_sainr_wang2024_media}; and two multi-modal methods: T1-conditioned ALPINE-A2 (ALPINE) \cite{ref_vyas2025_alpine_workshop}, Multi-contrast INR (MC-INR) \cite{ref_mcinr_mcginnis2023}. SA-INR requires separate training set, while other methods don't.

\subsubsection{Quantitative Results}

In MAE, PSNR and SSIM, Our method achieves the best performance on UKBB across all degradation factors, with a widening margin under severe downsampling (e.g., $\times7$, typical in clinical protocols), indicating improved robustness under sparse thick-slice supervision (Table~\ref{tab:ukbb}). Tumor preservation in the GBM cohort was assessed using nn-UNet Dice scores \cite{isensee2021nnunet}, where our model maintains stable lesion structure while modality-agnostic baselines often distort pathology and may even underperform linear interpolation. The method also generalizes to rare modalities with non-integral scaling ratios in the ABCD dataset (Table~\ref{tab:gbm}). Ablation Study shown in Table~\ref{tab:abla} supported the value of (1) topology perservation in stablizing gaussian representation; (2) latent appearance fitting for refined edge consistency; and (3) LR-consistent fusion for suppressing high-frequency gaussian blobs and artifacts. 

\begin{table*}[t]
\renewcommand{\arraystretch}{1.15}
\centering
\small
\fontsize{8}{9}\selectfont
\setlength{\tabcolsep}{3pt}
\caption{Quantitative comparison results on UK Biobank (Flair) for through-plane super-resolution at $\times3$, $\times5$, and $\times7$. Metrics are calculated inside brain mask. \textbf{Bold} indicates best results. \underline{Underline} indicates second-best results. Numbers in brackets mean half-width Confidence Interval (CI).}
\label{tab:ukbb}
\begin{tabular}{l|ccc|ccc|ccc}
\noalign{\hrule height 0.8pt}
& \multicolumn{3}{c|}{UKBB $\times3$} 
& \multicolumn{3}{c|}{UKBB $\times5$} 
& \multicolumn{3}{c}{UKBB $\times7$} \\
\hline
Method 
& MAE$\downarrow$ & SSIM$\uparrow$ & PSNR$\uparrow$ 
& MAE$\downarrow$ & SSIM$\uparrow$ & PSNR$\uparrow$ 
& MAE$\downarrow$ & SSIM$\uparrow$ & PSNR$\uparrow$  \\
\noalign{\hrule height 0.8pt}
Interp
& \begin{tabular}[c]{@{}c@{}}48.60\\(1.33)\end{tabular}
& \begin{tabular}[c]{@{}c@{}}0.8697\\(0.0014)\end{tabular}
& \begin{tabular}[c]{@{}c@{}}25.11\\(0.20)\end{tabular}
& \begin{tabular}[c]{@{}c@{}}65.94\\(1.81)\end{tabular}
& \begin{tabular}[c]{@{}c@{}}0.7866\\(0.0021)\end{tabular}
& \begin{tabular}[c]{@{}c@{}}22.64\\(0.20)\end{tabular}
& \begin{tabular}[c]{@{}c@{}}77.09\\(2.26)\end{tabular}
& \begin{tabular}[c]{@{}c@{}}0.7373\\(0.0026)\end{tabular}
& \begin{tabular}[c]{@{}c@{}}21.38\\(0.20)\end{tabular} \\[1.2ex]

Cubic
& \begin{tabular}[c]{@{}c@{}}46.78\\(1.28)\end{tabular}
& \begin{tabular}[c]{@{}c@{}}\underline{0.8793}\\(\underline{0.0013})\end{tabular}
& \begin{tabular}[c]{@{}c@{}}25.35\\(0.20)\end{tabular}
& \begin{tabular}[c]{@{}c@{}}65.22\\(1.78)\end{tabular}
& \begin{tabular}[c]{@{}c@{}}0.7919\\(0.0021)\end{tabular}
& \begin{tabular}[c]{@{}c@{}}22.66\\(0.20)\end{tabular}
& \begin{tabular}[c]{@{}c@{}}76.71\\(2.25)\end{tabular}
& \begin{tabular}[c]{@{}c@{}}\underline{0.7394}\\(\underline{0.0025})\end{tabular}
& \begin{tabular}[c]{@{}c@{}}21.34\\(0.20)\end{tabular} \\[1.2ex]

MC-INR
& \begin{tabular}[c]{@{}c@{}}53.44\\(1.48)\end{tabular}
& \begin{tabular}[c]{@{}c@{}}0.8729\\(0.0056)\end{tabular}
& \begin{tabular}[c]{@{}c@{}}24.69\\(0.28)\end{tabular}
& \begin{tabular}[c]{@{}c@{}}72.92\\(2.02)\end{tabular}
& \begin{tabular}[c]{@{}c@{}}0.7857\\(0.0058)\end{tabular}
& \begin{tabular}[c]{@{}c@{}}22.07\\(0.23)\end{tabular}
& \begin{tabular}[c]{@{}c@{}}86.25\\(2.44)\end{tabular}
& \begin{tabular}[c]{@{}c@{}}0.7284\\(0.0068)\end{tabular}
& \begin{tabular}[c]{@{}c@{}}20.71\\(0.25)\end{tabular} \\[1.2ex]

SA-INR
& \begin{tabular}[c]{@{}c@{}}46.95\\(1.27)\end{tabular}
& \begin{tabular}[c]{@{}c@{}}0.8758\\(0.0013)\end{tabular}
& \begin{tabular}[c]{@{}c@{}}25.17\\(0.20)\end{tabular}
& \begin{tabular}[c]{@{}c@{}}67.23\\(1.86)\end{tabular}
& \begin{tabular}[c]{@{}c@{}}0.7703\\(0.0026)\end{tabular}
& \begin{tabular}[c]{@{}c@{}}22.13\\(0.20)\end{tabular}
& \begin{tabular}[c]{@{}c@{}}80.15\\(2.35)\end{tabular}
& \begin{tabular}[c]{@{}c@{}}0.7037\\(0.0031)\end{tabular}
& \begin{tabular}[c]{@{}c@{}}20.69\\(0.21)\end{tabular} \\[1.2ex]

ALPINE
& \begin{tabular}[c]{@{}c@{}}239.35\\(7.53)\end{tabular}
& \begin{tabular}[c]{@{}c@{}}0.4652\\(0.0071)\end{tabular}
& \begin{tabular}[c]{@{}c@{}}13.17\\(0.24)\end{tabular}
& \begin{tabular}[c]{@{}c@{}}236.06\\(7.49)\end{tabular}
& \begin{tabular}[c]{@{}c@{}}0.4663\\(0.0073)\end{tabular}
& \begin{tabular}[c]{@{}c@{}}13.25\\(0.24)\end{tabular}
& \begin{tabular}[c]{@{}c@{}}232.46\\(7.72)\end{tabular}
& \begin{tabular}[c]{@{}c@{}}0.4662\\(0.0077)\end{tabular}
& \begin{tabular}[c]{@{}c@{}}13.31\\(0.25)\end{tabular} \\[1.2ex]

MedGS
& \begin{tabular}[c]{@{}c@{}}\underline{46.18}\\(\underline{1.47})\end{tabular}
& \begin{tabular}[c]{@{}c@{}}0.8783\\(0.0032)\end{tabular}
& \begin{tabular}[c]{@{}c@{}}\underline{25.68}\\(\underline{0.24})\end{tabular}
& \begin{tabular}[c]{@{}c@{}}\underline{62.57}\\(\underline{2.08})\end{tabular}
& \begin{tabular}[c]{@{}c@{}}\underline{0.7924}\\(\underline{0.0053})\end{tabular}
& \begin{tabular}[c]{@{}c@{}}\underline{23.17}\\(\underline{0.25})\end{tabular}
& \begin{tabular}[c]{@{}c@{}}\underline{72.79}\\(\underline{2.47})\end{tabular}
& \begin{tabular}[c]{@{}c@{}}0.7362\\(0.0055)\end{tabular}
& \begin{tabular}[c]{@{}c@{}}\underline{21.91}\\(\underline{0.25})\end{tabular} \\[1.2ex]

\textbf{Ours}
& \begin{tabular}[c]{@{}c@{}}\textbf{40.16}\\(\textbf{1.14})\end{tabular}
& \begin{tabular}[c]{@{}c@{}}\textbf{0.9047}\\(\textbf{0.0022})\end{tabular}
& \begin{tabular}[c]{@{}c@{}}\textbf{27.01}\\(\textbf{0.21})\end{tabular}
& \begin{tabular}[c]{@{}c@{}}\textbf{52.21}\\(\textbf{1.56})\end{tabular}
& \begin{tabular}[c]{@{}c@{}}\textbf{0.8464}\\(\textbf{0.0041})\end{tabular}
& \begin{tabular}[c]{@{}c@{}}\textbf{24.90}\\(\textbf{0.22})\end{tabular}
& \begin{tabular}[c]{@{}c@{}}\textbf{60.90}\\(\textbf{2.01})\end{tabular}
& \begin{tabular}[c]{@{}c@{}}\textbf{0.8004}\\(\textbf{0.0059})\end{tabular}
& \begin{tabular}[c]{@{}c@{}}\textbf{23.69}\\(\textbf{0.25})\end{tabular} \\[1.2ex]
\noalign{\hrule height 0.8pt}
\end{tabular}
\end{table*}

\begin{table*}[t]
\renewcommand{\arraystretch}{1.1}
\centering
\small
\fontsize{8}{9}\selectfont
\setlength{\tabcolsep}{4pt}
\caption{Quantitative Results on GBM (T2w, Flair) and ABCD (DWI, ASL). Metrics are calculated inside \textbf{Tumor Mask} for GBM and brain mask for ABCD. \textbf{DSC}: Dice score between groundtruth and predicted tumor mask using nn-UNet on reconstructions. \textbf{Bold} indicates best results. \underline{Underline} indicates second-best results.}
\label{tab:gbm}
\begin{tabular}{l|ccc|cc|cc|cc}
\noalign{\hrule height 0.8pt}
& \multicolumn{3}{c|}{GBM T2w $\times7$}
& \multicolumn{2}{c|}{GBM Flair $\times7$}
& \multicolumn{2}{c|}{ABCD DWI $\times3$}
& \multicolumn{2}{c}{ABCD ASL $\times3$} \\
\hline
Method
& MAE$\downarrow$ & SSIM$\uparrow$ & PSNR$\uparrow$
& MAE$\downarrow$ & DSC$\uparrow$
& MAE$\downarrow$ & SSIM$\uparrow$
& MAE$\downarrow$ & SSIM$\uparrow$ \\
\noalign{\hrule height 0.8pt}

Interp
& 149.15 & 0.8273 & 20.26
& 44.12 & 0.6572
& 9.65 & 0.9792
& 7.11 & 0.9802 \\

Cubic
& \underline{107.42} & 0.9055 & 21.63
& \underline{29.27} & 0.7247
& \textbf{5.78} & \textbf{0.9863}
& \underline{4.63} & \underline{0.9884} \\

SA-INR
& 108.97 & \underline{0.9075} & 21.35
& 29.37 & 0.7243
& \underline{5.85} & \underline{0.9863}
& 4.74 & 0.9877 \\

MedGS
& 109.39 & 0.8782 & \underline{21.79}
& 31.38 & \textbf{0.7714}
& 8.07 & 0.9824
& 4.89 & 0.9869 \\

\textbf{Ours}
& \textbf{86.75} & \textbf{0.9133} & \textbf{23.63}
& \textbf{27.36} & \underline{0.7458}
& 7.22 & 0.9859
& \textbf{4.60} & \textbf{0.9887} \\
\noalign{\hrule height 0.8pt}
\end{tabular}
\end{table*}

\begin{table*}[t]
\renewcommand{\arraystretch}{1.1}
\centering
\small
\fontsize{8}{9}\selectfont
\setlength{\tabcolsep}{3pt}
\caption{Ablation Study Results. \textbf{T1-guided}: 2-stage fitting with free appearance fitting and only reconstruction loss; \textbf{+Topology}: T1-guided + Persist Loss; \textbf{+Latent}: T1-guided + Latent Appearance Fitting; \textbf{+LR-Fusion}: T1-guided + LR Consistent Fusion; \textbf{+All}: With all innovations.}
\label{tab:abla}
\begin{tabular}{l|ccc|ccc|ccc}
\noalign{\hrule height 0.8pt}
& \multicolumn{3}{c|}{UKBB $\times3$} 
& \multicolumn{3}{c|}{UKBB $\times5$} 
& \multicolumn{3}{c}{UKBB $\times7$} \\
\hline
Method 
& MAE$\downarrow$ & SSIM$\uparrow$ & PSNR$\uparrow$ 
& MAE$\downarrow$ & SSIM$\uparrow$ & PSNR$\uparrow$ 
& MAE$\downarrow$ & SSIM$\uparrow$ & PSNR$\uparrow$  \\
\noalign{\hrule height 0.8pt}
T1-guided
& 50.74 & 0.8646 & 24.87
& 63.12 & 0.8056 & 23.09
& 72.80 & 0.7565 & 21.98 \\
+Topology
& 50.74 & 0.8646 & 24.87
& 63.06 & 0.8057 & 23.10
& 72.74 & 0.7566 & 21.99 \\
+Latent
& 50.25 & 0.8687 & 24.92
& 62.71 & 0.8092 & 23.13
& 75.71 & 0.7585 & 21.92 \\
+LR-Fusion
& 40.50 & 0.9028 & 26.96
& 52.53 & 0.8443 & 24.86
& 61.48 & 0.7972 & 23.62 \\
+All
& 40.16 & 0.9047 & 27.01
& 52.21 & 0.8464 & 24.90
& 61.19 & 0.7998 & 23.66 \\
\noalign{\hrule height 0.8pt}
\end{tabular}
\end{table*}

\subsubsection{Qualitative Results}

\begin{figure}[h]
\centering
\includegraphics[width=\textwidth]{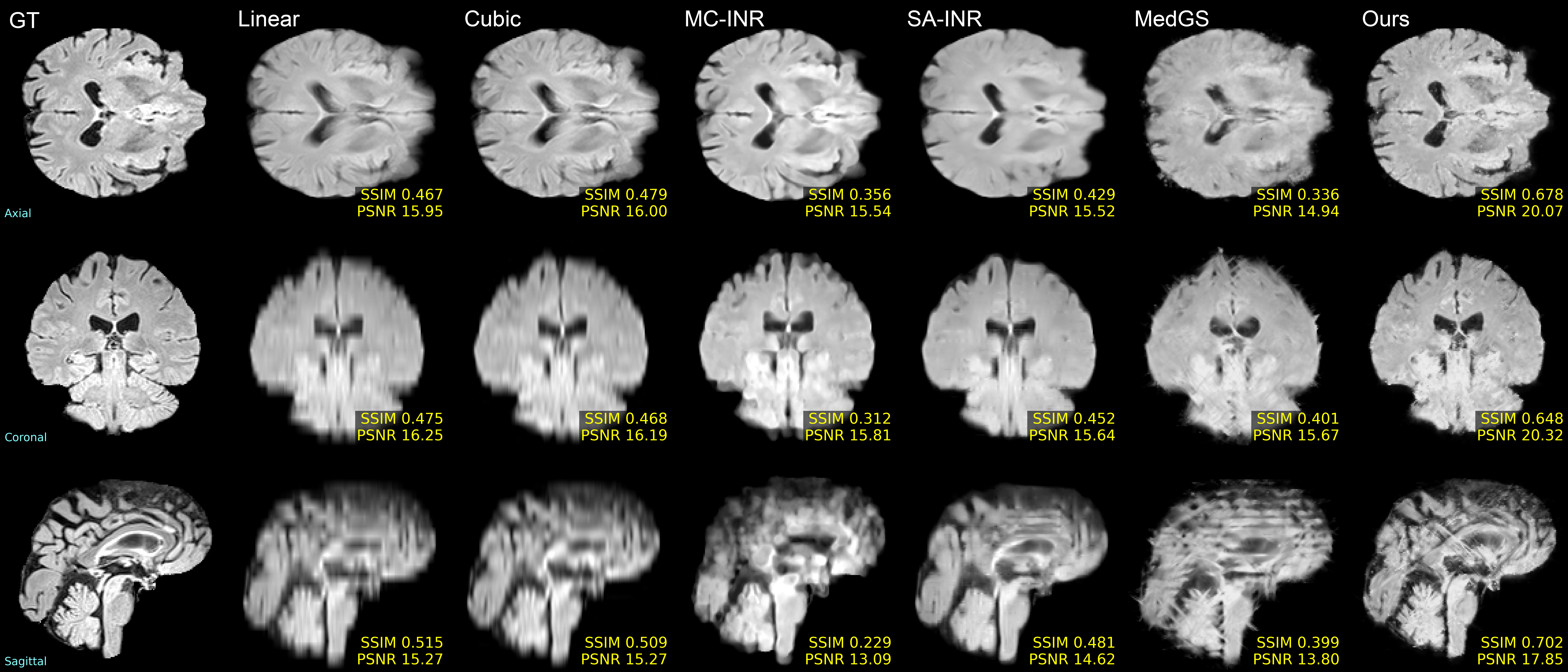}
\caption{Axial, Coronal and Sagittal view of reconstructions results with comparative methods. 7x axial super-resolution from UKBB Flair. From left to right: Groundtruth, Linear, Cubic, MC-INR, SA-INR, MedGS, Ours. }
\label{fig2}
\end{figure}

\begin{figure}[ht]
\centering
\includegraphics[width=\textwidth]{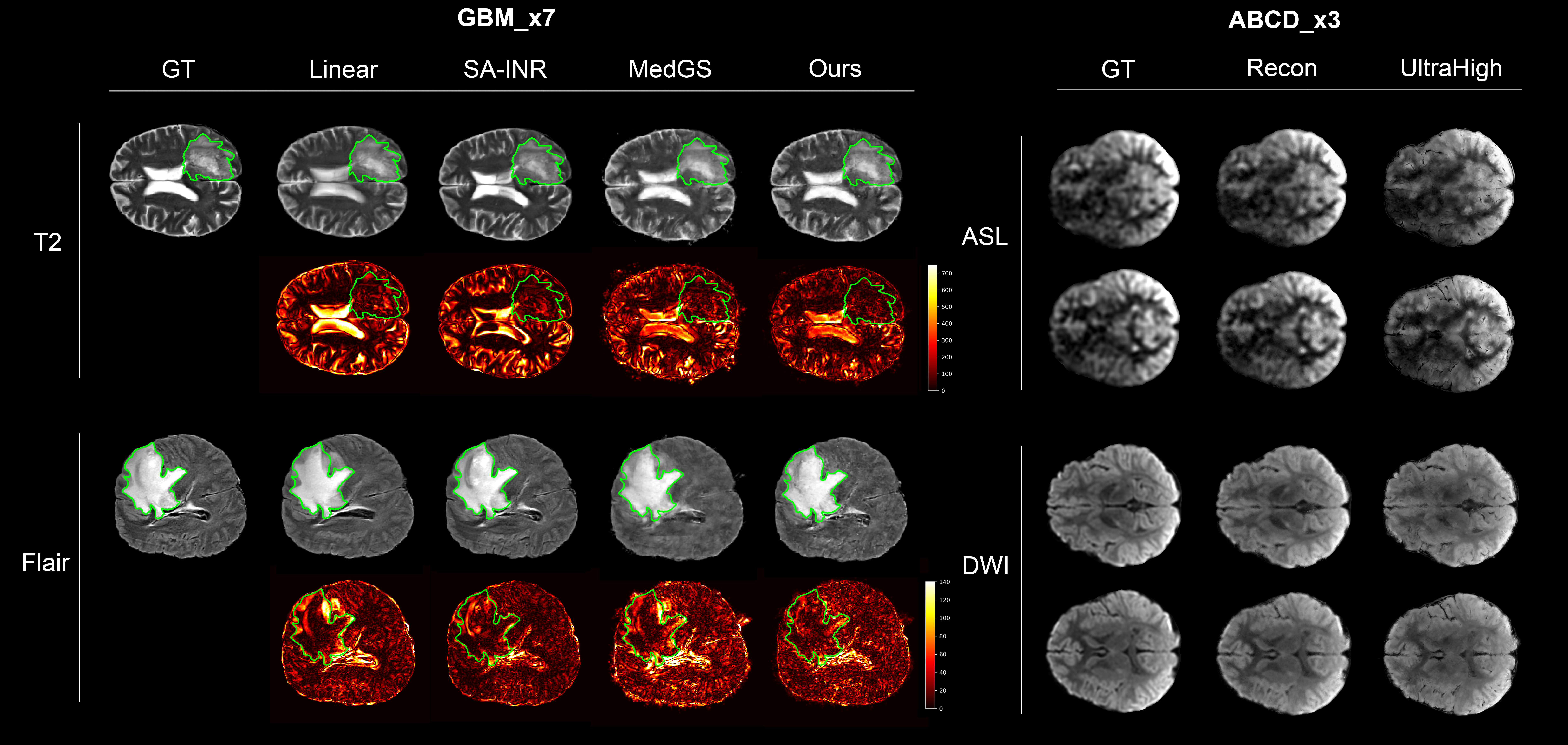}
\caption{Reconstruction results on GBM and ABCD datasets. GBM: green contour shows tumor regions, and error maps are given for reference. ABCD: UltraHigh denotes reconstructions generated at a sampling density matched to isotropic T1 resolution.}
\label{fig3}
\end{figure}

High-resolution reconstructions on axial, coronal and sagittal view proved our model managed to learn a continous high-resolution representation (Fig.~\ref{fig2}). Visualization on GBM dataset showed clear tumor boundary and eliminated halluciation introduced via interpolation, stressing clinical feasibility (Fig.~\ref{fig3} left). We also applied the Gaussian scaffold on T1 spacing and got ultra-high-resolution DWI and ASL scans in the ABCD dataset (Fig.~\ref{fig3} right).

\section{Discussion and Conclusion}

Our shared Gaussian framework achieves state-of-the-art through-plane reconstruction across multiple modalities and degradation levels without external training. The explicit shared geometric space restores in-plane structural detail and enables zero-shot adaptation of volume appearance.

Although geometry preservation is stable under anisotropic sampling, the framework has not been evaluated under motion corruption or anatomical deformation. Also the current formulation has not yet been extended to population-level atlas modeling. Expanding the shared scaffold toward atlas construction and large-scale harmonization is a natural next step.

Our geometry-aware design highlights both the value of structural priors as a lightweight yet powerful anchor, and the potential of explicit representation in medical image reconstruction. This paradigm enables both prospective protocol design and retrospective dataset reconciliation. Thanks to the flexibility and interpretability of gaussians, such explicit geometry-guided representations may provide a scalable foundation for cross-modal registration and mapping, longitudinal analysis, and image-guided intervention beyond super-resolution.

\bibliographystyle{splncs04}
\bibliography{mybibliography}
\end{document}